# On the physical interpretation of some types of three-dimensional harmonic mappings

**Andrey Petrin**

*Joint Institute for High Temperatures, Russian Academy of Science, Izhorskaya 13, st. 2, Moscow, 125412 Russia*

E-mail: a_petrin@mail.ru

**Abstract**

The development of the theory of three-dimensional harmonic mappings is considered. The new classes of mappings that generate three-dimensional harmonic functions are introduced. The physical interpretation of these mappings is applied to electrostatics problems. It is found that these mappings locally conserve electric charge of the equipotential surfaces. To confirm the correctness of the theory it is shown that by using the proposed mappings the electric field in two known electrostatic problems can be found.

**1. Introduction**

The theory of conformal mappings of two-dimensional domains defined by complex differentiable functions is used for various applications, including electrostatics and hydrodynamics [1, 2, 3]. A large class of plane problems with free boundaries could be investigated through the use of the methods of conformal mappings [4]. However, when trying to generalize these methods for three-dimensional problems there are serious problems. The difficulty and importance of three-dimensional physical problems have stimulated the search three-dimensional analogues of the mappings, based on which one could develop new methods for solving problems with free boundaries. The searches for such analogues have been conducted for a long time [5]. However, in three-dimensional space is not a sufficiently broad class of mappings that preserve all the basic properties of conformal mappings of plane regions, so we have to consider such mappings, which preserve only a portion of the properties of conformal mappings.

The most important class of three-dimensional mappings of domains is obtained by the generalization on three-dimensional case of hydrodynamic or electrostatic interpretation of the conformal mapping of plane regions. Systematically such approach was developed by M. A. Lavrent'ev. It was suggested the mappings defined in rectangular Cartesian coordinate systems. However, as will be shown below, the form of mappings, in general, depends on the choice of coordinate systems in which this mapping is given. The first mention of harmonic mappings defined in the orthogonal curvilinear coordinate systems, apparently, was given without rigorous proofs in a little-known work [6]. In this study, these results are significantly refined, expanded and given rigorous proofs of harmony of the proposed mappings. In addition, to confirm the correctness of the theory it is shown that the solution of two well-known problems of calculating the electrostatic field of a charged metal disk and a charged thin metal needle of finite length may be found with the aid of the proposed harmonic mappings.

## 2. Determination of harmonic mappings

Consider a three-dimensional mapping of a region to another. Let $(x, y, z)$ and $(u, v, w)$ – systems of rectangular Cartesian coordinates in these regions (see Fig. 1). We assume that the orts vectors $\mathbf{i}, \mathbf{j}, \mathbf{k}$ of these coordinate systems are identical. Suppose that on this mapping is imposed condition (which will serve as an analog of the Cauchy-Riemann conditions of two-dimensional conformal mappings):

$$\mathbf{i} u_z + \mathbf{j} v_z + \mathbf{k} w_z = \begin{vmatrix} \mathbf{i} & \mathbf{j} & \mathbf{k} \\ u_x & v_x & w_x \\ u_y & v_y & w_y \end{vmatrix}. \qquad (1)$$

This vector condition is equivalent to the following three scalar equations

$$\begin{aligned} u_z &= v_x w_y - v_y w_x, \\ v_z &= u_y w_x - u_x w_y, \\ w_z &= u_x v_y - u_y v_x. \end{aligned} \qquad (2)$$

The condition (1) geometrically means the following. Consider a vector $\mathbf{i}dx$ in the space $(x, y, z)$. In the space $(u, v, w)$ this vector corresponds to the vector $(\mathbf{i}u_x + \mathbf{j}v_x + \mathbf{k}w_x)dx$.

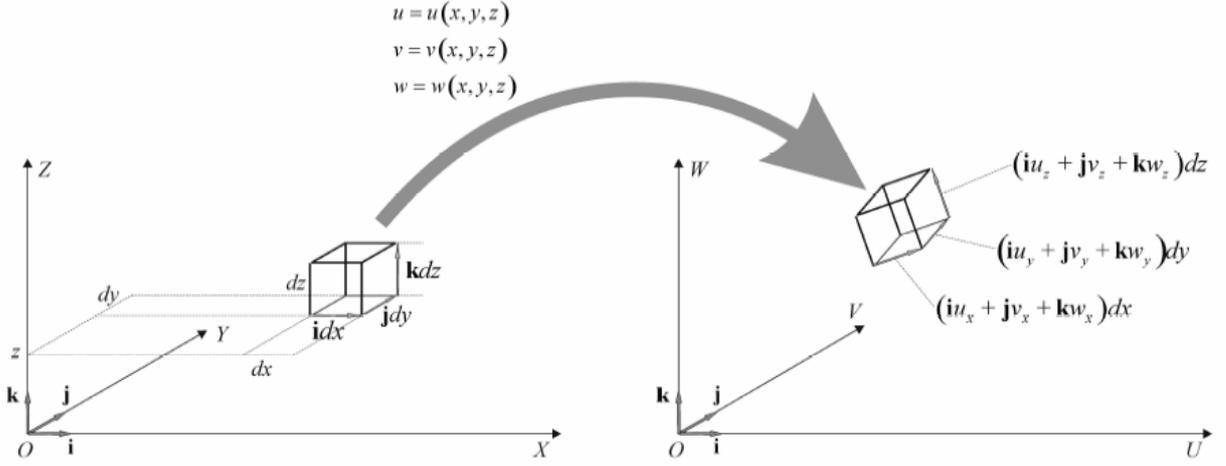

Fig. 1. Mapping of a region of space with the coordinates $(x, y, z)$ in a region of space with the coordinates $(u, v, w)$.

Similarly, the vector $\mathbf{j}dy$ corresponds to the vector $(\mathbf{i}u_y + \mathbf{j}v_y + \mathbf{k}w_y)dy$ and the vector $\mathbf{k}dz$ corresponds to the vector $(\mathbf{i}u_z + \mathbf{j}v_z + \mathbf{k}w_z)dz$. Then, condition (1) means that the last vector is the vector product of the first two, if $dz = dxdy$. That is

$$\mathbf{i}u_z + \mathbf{j}v_z + \mathbf{k}w_z = (\mathbf{i}u_x + \mathbf{j}v_x + \mathbf{k}w_x) \times (\mathbf{i}u_y + \mathbf{j}v_y + \mathbf{k}w_y).$$

From the properties of vector multiplication (the resulting vector is perpendicular to each of the factors) it follows that

$$u_z u_x + v_z v_x + w_z w_x = 0 \,;\quad u_z u_y + v_z v_y + w_z w_y = 0 \,.$$

Considering the functions $x = x(u, v, w)$, $y = y(u, v, w)$ and $z = z(u, v, w)$ we can write the equations:

$$\frac{\partial x}{\partial x} = x_u u_x + x_v v_x + x_w w_x = 1 \,;\quad \frac{\partial x}{\partial y} = x_u u_y + x_v v_y + x_w w_y = 0 \,;\quad \frac{\partial x}{\partial z} = x_u u_z + x_v v_z + x_w w_z = 0 \,;$$

$$\frac{\partial y}{\partial x} = y_u u_x + y_v v_x + y_w w_x = 0 \,;\quad \frac{\partial y}{\partial y} = y_u u_y + y_v v_y + y_w w_y = 1 \,;\quad \frac{\partial y}{\partial z} = y_u u_z + y_v v_z + y_w w_z = 0 \,;$$

$$\frac{\partial z}{\partial x} = z_u u_x + z_v v_x + z_w w_x = 0 \, ; \; \frac{\partial z}{\partial y} = z_u u_y + z_v v_y + z_w w_y = 0 \, ; \; \frac{\partial z}{\partial z} = z_u u_z + z_v v_z + z_w w_z = 1 \, .$$

In the form of a matrix equation the resulting equations can be written as:

$$\begin{bmatrix} u_x & v_x & w_x & 0 & 0 & 0 & 0 & 0 & 0 \\ u_y & v_y & w_y & 0 & 0 & 0 & 0 & 0 & 0 \\ u_z & v_z & w_z & 0 & 0 & 0 & 0 & 0 & 0 \\ 0 & 0 & 0 & u_x & v_x & w_x & 0 & 0 & 0 \\ 0 & 0 & 0 & u_y & v_y & w_y & 0 & 0 & 0 \\ 0 & 0 & 0 & u_z & v_z & w_z & 0 & 0 & 0 \\ 0 & 0 & 0 & 0 & 0 & 0 & u_x & v_x & w_x \\ 0 & 0 & 0 & 0 & 0 & 0 & u_y & v_y & w_y \\ 0 & 0 & 0 & 0 & 0 & 0 & u_z & v_z & w_z \end{bmatrix} \times \begin{bmatrix} x_u \\ x_v \\ x_w \\ y_u \\ y_v \\ y_w \\ z_u \\ z_v \\ z_w \end{bmatrix} = \begin{bmatrix} 1 \\ 0 \\ 0 \\ 0 \\ 1 \\ 0 \\ 0 \\ 0 \\ 1 \end{bmatrix}. \quad (3)$$

The determinant $J$ of the matrix in (3) is equal to $J = D^3$, where

$$D = \begin{vmatrix} u_z & v_z & w_z \\ u_x & v_x & w_x \\ u_y & v_y & w_y \end{vmatrix} = u_z \left( v_x w_y - v_y w_x \right) + v_z \left( u_y w_x - u_x w_y \right) + w_z \left( u_x v_y - u_y v_x \right).$$

Taking into account (2), we obtain:

$$D = \left( u_z \right)^2 + \left( v_z \right)^2 + \left( w_z \right)^2 . \quad (4)$$

Assuming that $D \ne 0$ from the system of equations (3) it follows

$$z_u = D^{-1} \left( v_x w_y - v_y w_x \right), \; z_v = D^{-1} \left( u_y w_x - u_x w_y \right), \; z_w = D^{-1} \left( u_x v_y - u_y v_x \right); \quad (5)$$

$$x_u = D^{-1} \left( v_y w_z - v_z w_y \right), \; x_v = D^{-1} \left( u_z w_y - u_y w_z \right), \; x_w = D^{-1} \left( u_y v_z - u_z v_y \right); \quad (6)$$

$$y_u = D^{-1} \left( v_z w_x - v_x w_z \right), \; y_v = D^{-1} \left( u_x w_z - u_z w_x \right), \; y_w = D^{-1} \left( u_z v_x - u_x v_z \right). \quad (7)$$

$$u_z = v_x w_y - v_y w_x,$$
$$v_z = u_y w_x - u_x w_y,$$
$$w_z = u_x v_y - u_y v_x.$$

By substituting (5)-(7) into (2), we obtain

$$u_z = D z_u, \quad (8)$$

$$v_z = D z_v, \quad (9)$$

$$w_z = D z_w. \quad (10)$$

By substituting (8)-(10) into (6) and (7), we find

$$x_u = D^{-1}\left(v_y Dz_w - Dz_v w_y\right), \ x_v = D^{-1}\left(Dz_u w_y - u_y Dz_w\right), \ x_w = D^{-1}\left(u_y Dz_v - Dz_u v_y\right);$$

$$y_u = D^{-1}\left(Dz_v w_x - v_x Dz_w\right), \ y_v = D^{-1}\left(u_x Dz_w - Dz_u w_x\right), \ y_w = D^{-1}\left(Dz_u v_x - u_x Dz_v\right).$$

Therefore, we obtain

$$x_u y_w - y_u x_w =$$
$$= z_u z_w \left(v_x v_y - v_x v_y\right) + z_u z_v \left(v_y w_x - v_x w_y\right) + z_v z_w \left(v_x u_y - u_x v_y\right) + z_v z_v \left(u_x w_y - w_x u_y\right) =$$
$$= z_v \left\{z_u \left(v_y w_x - v_x w_y\right) + z_w \left(v_x u_y - u_x v_y\right) + z_v \left(u_x w_y - w_x u_y\right)\right\}.$$

Taking into account (8)-(10), we find

$$x_u y_w - y_u x_w = \frac{z_v}{D}\left\{u_z \left(v_y w_x - v_x w_y\right) + w_z \left(v_x u_y - u_x v_y\right) + v_z \left(u_x w_y - w_x u_y\right)\right\}.$$

The expression in the parentheses is equal to $-D$, therefore $z_v = y_u x_w - x_u y_w$. Similarly, we may find $z_w = x_u y_v - x_v y_u$ and $z_u = x_v y_w - x_w y_v$.

The result is the formulas for the inverse transformation:

$$\begin{aligned} z_u &= x_v y_w - x_w y_v, \\ z_v &= y_u x_w - x_u y_w, \\ z_w &= x_u y_v - x_v y_u. \end{aligned} \qquad (11)$$

Note that conditions (11) differ from the conditions (2) for direct transformation.

Let prove that the function $z = z(u,v,w)$ satisfies the Laplace equation:

$$z_{uu} + z_{vv} + z_{ww} = 0. \qquad (12)$$

Indeed,

$$\begin{aligned} z_{uu} + z_{vv} + z_{ww} &= \frac{\partial}{\partial u}\left(x_v y_w - x_w y_v\right) + \frac{\partial}{\partial v}\left(y_u x_w - x_u y_w\right) + \frac{\partial}{\partial w}\left(x_u y_v - x_v y_u\right) = \\ &= x_{vu} y_w + x_v y_{wu} - x_{wu} y_v - x_w y_{vu} + y_{uv} x_w + y_u x_{wv} - x_{uv} y_w - x_u y_{wv} + x_{uw} y_v + x_u y_{vw} - \\ &\quad - x_{vw} y_u - x_v y_{uw} = \\ &= \left\{x_{vu} y_w - x_{uv} y_w\right\} + \left\{x_v y_{wu} - x_v y_{uw}\right\} + \left\{x_{uw} y_v - x_{wu} y_v\right\} + \left\{y_{uv} x_w - x_w y_{vu}\right\} + \\ &\quad + \left\{y_u x_{wv} - x_{vw} y_u\right\} + \left\{x_u y_{vw} - x_u y_{wv}\right\} = 0, \end{aligned} \qquad (13)$$

in view of the equality of cross derivatives in the brackets.

Eq.(11) known as the equations defining the harmonic on M. A. Lavrent'ev mapping [7].

## 3. The physical sense of harmonic mapping

Imagine that in the space $(x, y, z)$ there is a uniform electric field with potential $\psi = z$. Then the planes perpendicular to the axis $OZ$ are equipotential planes.

Consider two near planes $a$ and $b$ determined the values of potential $z$ and $z + dz$ (see Fig. 2), respectively. For harmonic mapping (1) these two planes transform to the equipotential surfaces $A$ and $B$. Consider the volume element $dxdydz$ that distinguishes on an equipotential plane $a$ the base area $ds = dxdy$. The image of this volume element in the space $(u, v, w)$ is the volume element with height $|\mathbf{i}u_z + \mathbf{j}v_z + \mathbf{k}w_z|dz$ and the base area $dS = |(\mathbf{i}u_x + \mathbf{j}v_x + \mathbf{k}w_x) \times (\mathbf{i}u_y + \mathbf{j}v_y + \mathbf{k}w_y)|ds$. In the previous section it was proved that the function $z = z(u, v, w)$ satisfies the Laplace equation, i.e. is a potential. Therefore, the surfaces $A$ and $B$ which are images of planes $a$ and $b$ have the same potential values $z$ and $z + dz$. If $dz = ds$, then the condition (1) means that

$$|\mathbf{i}u_z + \mathbf{j}v_z + \mathbf{k}w_z|dz = dS = |(\mathbf{i}u_x + \mathbf{j}v_x + \mathbf{k}w_x) \times (\mathbf{i}u_y + \mathbf{j}v_y + \mathbf{k}w_y)|ds.$$

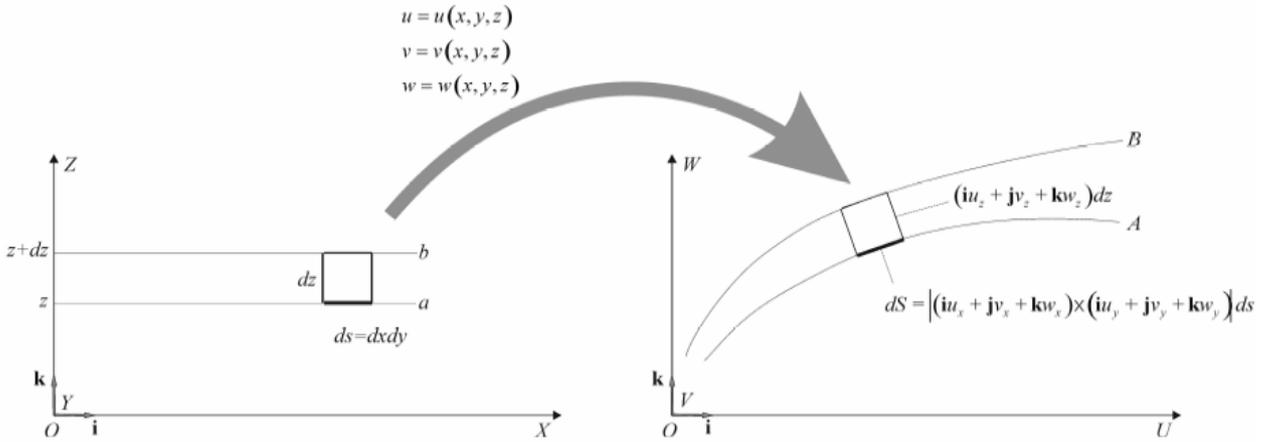

Fig. 2. Mapping of the volume element, located between close equipotentials.

Suppose that the equipotential plane $a$ is replaced by a metal infinitely thin plane. With this change, the electric field distribution does not change. Then, on the surface of the metal the surface electric charge density is $\sigma = -\varepsilon_0$, where $\varepsilon_0$ is the vacuum electric constant. The electric charge of $ds$ is equal to $dq = \sigma ds = -\varepsilon_0 ds$. In the space of images, it is also possible to replace the equipotential surface $A$ by infinitely thin layer of metal. Then the surface charge density on the metal is equal to

$$\Sigma = -\varepsilon_0 \frac{d\psi}{|\mathbf{i}u_z + \mathbf{j}v_z + \mathbf{k}w_z|dz} = -\frac{\varepsilon_0}{|\mathbf{i}u_z + \mathbf{j}v_z + \mathbf{k}w_z|}.$$

The electric charge of the surface $dS$ ($dS$ is the image of $ds$) is equal to

$$dQ = \Sigma dS = -\frac{\varepsilon_0}{|\mathbf{i}u_z + \mathbf{j}v_z + \mathbf{k}w_z|}\left|(\mathbf{i}u_x + \mathbf{j}v_x + \mathbf{k}w_x)\times(\mathbf{i}u_y + \mathbf{j}v_y + \mathbf{k}w_y)\right|ds.$$

We can see, that if the conditions (1) are fulfilled then $dq = dQ$. This means that under this mapping the charge of the element base of the preimage is equal to the charge of the element base of the image. That is, *the charge of an element under the mapping does not change.* In addition, *the electric field under the mapping is inversely proportional to the change in the area of the base of the element.*

As will be shown below, based on established law of conservation of charge it is possible to generalize (1) and obtain other types of harmonic mappings.

## 4. Harmonic mapping from the region with a cylindrical coordinate system into the region with its cylindrical coordinate system

Consider in the space a uniform electric field with a potential $\psi = z$ depended only on the coordinate $z$ of cylindrical coordinate system $(r,\varphi,z)$. Introduce the orts of this coordinate system $\mathbf{i}_r$, $\mathbf{i}_\varphi$ and $\mathbf{i}_z$. Then, as above, the planes perpendicular to the axis $OZ$ are equipotential planes. Consider two similar planes $a$ and $b$ determined by the values of potential $z$ and $z+dz$, respectively. The harmonic mapping transfers these two planes into the equipotential surfaces $A$ and $B$. Consider the volume element $rdrd\varphi dz$ on the equipotential plane $a$ with the base area $ds = rdrd\varphi$. Assume that $dz = ds$. The image of this volume element in the space of images (with a cylindrical coordinate system $(R,\Phi,Z)$ and coordinate orts $\mathbf{e}_R$, $\mathbf{e}_\Phi$ and $\mathbf{e}_Z$) is the element of height $|\mathbf{e}_R R_z + \mathbf{e}_\Phi R\Phi_z + \mathbf{e}_Z Z_z|dz$ and of base area

$$dS = \left|(\mathbf{e}_R R_r + \mathbf{e}_\Phi R\Phi_r + \mathbf{e}_Z Z_r)\times(\mathbf{e}_R R_\varphi + \mathbf{e}_\Phi R\Phi_\varphi + \mathbf{e}_Z Z_\varphi)\right|\frac{ds}{r}.$$

Conditions similar to (1) can be written as [7]:

$$\mathbf{e}_R R_z + \mathbf{e}_\Phi R\Phi_z + \mathbf{e}_Z Z_z = \frac{1}{r} \begin{vmatrix} \mathbf{e}_R & \mathbf{e}_\Phi & \mathbf{e}_Z \\ R_r & R\Phi_r & Z_r \\ R_\varphi & R\Phi_\varphi & Z_\varphi \end{vmatrix}. \tag{14}$$

Then, instead of (2) we obtain the equations

$$\begin{aligned} rR_z &= R\Phi_r Z_\varphi - R\Phi_\varphi Z_r, \\ rR\Phi_z &= R_\varphi Z_r - R_r Z_\varphi, \\ rZ_z &= R_r R\Phi_\varphi - R_\varphi R\Phi_r. \end{aligned} \tag{15}$$

Analog of matrix equation (3) in the case of cylindrical coordinates is the matrix equation:

$$\begin{bmatrix} R_r & \Phi_r & Z_r & 0 & 0 & 0 & 0 & 0 & 0 \\ R_\varphi & \Phi_\varphi & Z_\varphi & 0 & 0 & 0 & 0 & 0 & 0 \\ R_z & \Phi_z & Z_z & 0 & 0 & 0 & 0 & 0 & 0 \\ 0 & 0 & 0 & R_r & \Phi_r & Z_r & 0 & 0 & 0 \\ 0 & 0 & 0 & R_\varphi & \Phi_\varphi & Z_\varphi & 0 & 0 & 0 \\ 0 & 0 & 0 & R_z & \Phi_z & Z_z & 0 & 0 & 0 \\ 0 & 0 & 0 & 0 & 0 & 0 & R_r & \Phi_r & Z_r \\ 0 & 0 & 0 & 0 & 0 & 0 & R_\varphi & \Phi_\varphi & Z_\varphi \\ 0 & 0 & 0 & 0 & 0 & 0 & R_z & \Phi_z & Z_z \end{bmatrix} \times \begin{bmatrix} z_R \\ z_\Phi \\ z_Z \\ r_R \\ r_\Phi \\ r_Z \\ \varphi_R \\ \varphi_\Phi \\ \varphi_Z \end{bmatrix} = \begin{bmatrix} 0 \\ 0 \\ 1 \\ 1 \\ 0 \\ 0 \\ 0 \\ 1 \\ 0 \end{bmatrix}. \tag{16}$$

Eq.(16) is obtained by differentiation of functions $r = r(R, \Phi, Z)$, $\varphi = \varphi(R, \Phi, Z)$ and $z = z(R, \Phi, Z)$.

Determinant of the matrix $J_C$ in (16) is equal to

$$J_C = \begin{vmatrix} R_r & \Phi_r & Z_r \\ R_\varphi & \Phi_\varphi & Z_\varphi \\ R_z & \Phi_z & Z_z \end{vmatrix} \times \begin{vmatrix} R_r & \Phi_r & Z_r \\ R_\varphi & \Phi_\varphi & Z_\varphi \\ R_z & \Phi_z & Z_z \end{vmatrix} \times \begin{vmatrix} R_r & \Phi_r & Z_r \\ R_\varphi & \Phi_\varphi & Z_\varphi \\ R_z & \Phi_z & Z_z \end{vmatrix} = \begin{vmatrix} R_r & \Phi_r & Z_r \\ R_\varphi & \Phi_\varphi & Z_\varphi \\ R_z & \Phi_z & Z_z \end{vmatrix}^3 = D_C^3,$$

where $D_C = \begin{vmatrix} R_r & \Phi_r & Z_r \\ R_\varphi & \Phi_\varphi & Z_\varphi \\ R_z & \Phi_z & Z_z \end{vmatrix} = R_z(\Phi_r Z_\varphi - \Phi_\varphi Z_r) + \Phi_z(R_\varphi Z_r - R_r Z_\varphi) + Z_z(R_r \Phi_\varphi - R_\varphi \Phi_r).$

Taking into account (15), we obtain:

$$D_C = \frac{r}{R}(R_z)^2 + rR(\Phi_z)^2 + \frac{r}{R}(Z_z)^2 = \frac{r}{R}\left\{(R_z)^2 + (R\Phi_z)^2 + (Z_z)^2\right\}. \tag{17}$$

Assuming that $D_C \neq 0$, from the system of equations (16) it follows

$$z_R = D_C^{-1}(\Phi_r Z_\varphi - \Phi_\varphi Z_r), \; z_\Phi = D_C^{-1}(R_\varphi Z_r - R_r Z_\varphi), \; z_Z = D_C^{-1}(R_r \Phi_\varphi - R_\varphi \Phi_r); \tag{18}$$

$$r_R = D_C^{-1}\left(\Phi_\varphi Z_z - \Phi_z Z_\varphi\right),\ r_\Phi = D_C^{-1}\left(R_z Z_\varphi - R_\varphi Z_z\right),\ r_Z = D_C^{-1}\left(R_\varphi \Phi_z - R_z \Phi_\varphi\right); \quad (19)$$

$$\varphi_R = D_C^{-1}\left(\Phi_z Z_r - \Phi_r Z_z\right),\ \varphi_\Phi = D_C^{-1}\left(R_r Z_z - R_z Z_r\right),\ \varphi_Z = D_C^{-1}\left(R_z \Phi_r - R_r \Phi_z\right). \quad (20)$$

By substituting (18)-(20) into (15) we obtain

$$R_z = \frac{R}{r} D_C z_R, \quad (21)$$

$$\Phi_z = \frac{1}{rR} D_C z_\Phi, \quad (22)$$

$$Z_z = \frac{R}{r} D_C z_Z. \quad (23)$$

By substituting (21)-(23) into (19) and (20) we obtain

$$r_R = \frac{1}{r}\left(\Phi_\varphi R z_Z - \frac{1}{R} z_\Phi Z_\varphi\right),\ r_\Phi = \frac{R}{r}\left(z_R Z_\varphi - R_\varphi z_Z\right),\ r_Z = \frac{1}{r}\left(R_\varphi \frac{1}{R} z_\Phi - R z_R \Phi_\varphi\right); \quad (24)$$

$$\varphi_R = \frac{1}{r}\left(\frac{1}{R} z_\Phi Z_r - \Phi_r R z_Z\right),\ \varphi_\Phi = \frac{R}{r}\left(R_r z_Z - z_R Z_r\right),\ \varphi_Z = \frac{1}{r}\left(R z_R \Phi_r - R_r \frac{1}{R} z_\Phi\right). \quad (25)$$

Whence, we find the expression

$$r_R \varphi_Z - \varphi_R r_Z =$$
$$= \frac{1}{r^2 R^2}\left\{z_\Phi z_\Phi\left(R_r Z_\varphi - Z_r R_\varphi\right) - z_\Phi z_R\left(Z_\varphi R^2 \Phi_r - R^2 \Phi_\varphi Z_r\right) - z_\Phi z_Z\left(R_r R^2 \Phi_\varphi - R^2 \Phi_r R_\varphi\right)\right\}.$$

Taking into account (21)-(23) we have

$$r^2 R^2 \left(r_R \varphi_Z - \varphi_R r_Z\right) =$$
$$= \frac{rR}{D_C} z_\Phi \left\{\Phi_z\left(R_r Z_\varphi - Z_r R_\varphi\right) + R_z\left(\Phi_\varphi Z_r - Z_\varphi \Phi_r\right) + Z_z\left(\Phi_r R_\varphi - R_r \Phi_\varphi\right)\right\} \quad (26)$$

The expression in braces is equal to $-D_C$, so $z_\Phi = rR\left(\varphi_R r_Z - r_R \varphi_Z\right)$. Similarly, we obtain the expressions $z_Z = \frac{r}{R}\left(r_R \varphi_\Phi - r_\Phi \varphi_R\right)$ and $z_R = \frac{r}{R}\left(r_\Phi \varphi_Z - r_Z \varphi_\Phi\right)$.

The result is a condition for the inverse transformation in cylindrical coordinate systems, similar to Cauchy-Riemann conditions:

$$z_R = \frac{r}{R}(r_\Phi \varphi_Z - r_Z \varphi_\Phi),$$
$$z_\Phi = rR(\varphi_R r_Z - r_R \varphi_Z), \quad (27)$$
$$z_Z = \frac{r}{R}(r_R \varphi_\Phi - r_\Phi \varphi_R).$$

In vector form (27) can be written as

$$\mathbf{e}_r z_R + \mathbf{e}_\varphi \frac{z_\Phi}{R^2} + \mathbf{e}_z z_Z = \frac{1}{R} \begin{vmatrix} \mathbf{e}_r & \mathbf{e}_\varphi & \mathbf{e}_z \\ r_R & r_\Phi & r_Z \\ r\varphi_R & r\varphi_\Phi & r\varphi_Z \end{vmatrix}$$

Prove that the function $z = z(R, \Phi, Z)$ satisfies the Laplace equation:

$$\Delta_{(R\Phi Z)} z = \frac{1}{R}\frac{\partial}{\partial R}\left(R\frac{\partial z}{\partial R}\right) + \frac{1}{R^2}\frac{\partial^2 z}{\partial \Phi^2} + \frac{\partial^2 z}{\partial Z^2} = 0.$$

Indeed, using (27) we obtain

$$\Delta_{(R\Phi Z)} z = \frac{1}{R}\frac{\partial}{\partial R}(Rz_R) + \frac{1}{R^2}\frac{\partial z_\Phi}{\partial \Phi} + \frac{\partial z_Z}{\partial Z} = \frac{1}{R}\frac{\partial}{\partial R}\left(R\frac{r}{R}(r_\Phi\varphi_Z - r_Z\varphi_\Phi)\right) + \frac{1}{R^2}\frac{\partial}{\partial \Phi}(rR(\varphi_R r_Z - r_R\varphi_Z)) +$$
$$+ \frac{\partial}{\partial Z}\left(\frac{r}{R}(r_R\varphi_\Phi - r_\Phi\varphi_R)\right) = \frac{1}{R}\left\{\frac{\partial}{\partial R}(r(r_\Phi\varphi_Z - r_Z\varphi_\Phi)) + \frac{\partial}{\partial \Phi}(r(\varphi_R r_Z - r_R\varphi_Z)) + \frac{\partial}{\partial Z}(r(r_R\varphi_\Phi - r_\Phi\varphi_R))\right\} =$$
$$= \frac{1}{R}\left\{r_R(r_\Phi\varphi_Z - r_Z\varphi_\Phi) + r_\Phi(\varphi_R r_Z - r_R\varphi_Z) + r_Z(r_R\varphi_\Phi - r_\Phi\varphi_R)\right\} +$$
$$+ \frac{r}{R}\left\{\frac{\partial}{\partial R}(r_\Phi\varphi_Z - r_Z\varphi_\Phi) + \frac{\partial}{\partial \Phi}(\varphi_R r_Z - r_R\varphi_Z) + \frac{\partial}{\partial Z}(r_R\varphi_\Phi - r_\Phi\varphi_R)\right\}.$$

The expression in the first braces is

$$r_R(r_\Phi\varphi_Z - r_Z\varphi_\Phi) + r_\Phi(\varphi_R r_Z - r_R\varphi_Z) + r_Z(r_R\varphi_\Phi - r_\Phi\varphi_R) =$$
$$= r_R r_\Phi \varphi_Z - r_R r_Z \varphi_\Phi + \varphi_R r_\Phi r_Z - r_\Phi r_R \varphi_Z + r_Z r_R \varphi_\Phi - r_Z r_\Phi \varphi_R =$$
$$= [r_R r_\Phi \varphi_Z - r_\Phi r_R \varphi_Z] + [r_Z r_R \varphi_\Phi - r_R r_Z \varphi_\Phi] + [\varphi_R r_\Phi r_Z - r_Z r_\Phi \varphi_R] \equiv 0.$$

The expression in the second braces (in view of the equality of cross derivatives) is

$$\frac{\partial}{\partial R}(r_\Phi\varphi_Z - r_Z\varphi_\Phi) + \frac{\partial}{\partial \Phi}(\varphi_R r_Z - r_R\varphi_Z) + \frac{\partial}{\partial Z}(r_R\varphi_\Phi - r_\Phi\varphi_R) =$$
$$= r_{R\Phi}\varphi_Z + r_\Phi\varphi_{RZ} - r_{RZ}\varphi_\Phi - r_Z\varphi_{R\Phi} + \varphi_{\Phi R} r_Z + \varphi_R r_{\Phi Z} - \varphi_{\Phi R}\varphi_Z - r_R\varphi_{\Phi Z} +$$
$$+ r_{ZR}\varphi_\Phi - r_Z\varphi_{\Phi R} + r_R\varphi_{Z\Phi} - r_\Phi\varphi_{ZR} =$$
$$= [r_{R\Phi}\varphi_Z - r_{\Phi R}\varphi_Z] + [r_\Phi\varphi_{RZ} - r_\Phi\varphi_{ZR}] + [r_{ZR}\varphi_\Phi - r_{RZ}\varphi_\Phi] + [\varphi_{\Phi R} r_Z - r_Z\varphi_{R\Phi}] +$$
$$+ [\varphi_R r_{\Phi Z} - r_{Z\Phi}\varphi_R] + [r_R\varphi_{Z\Phi} - r_R\varphi_{\Phi Z}] = 0.$$

Therefore $\Delta_{(R\Phi Z)} z = 0$ is proven.

Note that by analogy one can find harmonic mappings from region with a cylindrical coordinate system into the region with a spherical coordinate system, or from the region with Cartesian coordinate system into the region with a cylindrical (or spherical coordinate system). Further generalization of the proposed method to obtain formulas of harmonic mappings on curvilinear orthogonal coordinates in the regions will be given below in Section 6.

## 5. A particular case of harmonic mapping

An important particular case of transformations (15) is obtained when $\Phi = \varphi$, $R = R(r,z)$ and $Z = Z(r,z)$:

$$rR_z = -RZ_r,$$
$$rZ_z = RR_r. \quad (28)$$

Not pretending to a complete analysis of the system of equations (28), we consider as an example two physically important partial solutions of (28).

a). Let seek a solution in the form $Z = A(r)tg[\alpha(z+F)]$, $R = B(r)/\cos[\alpha(z+F)]$, where $\alpha$ and $F$ are constants. Then it is easy to find a solution of (28):

$$R = \pm\frac{1}{\cos[\alpha(z+F)]}\sqrt{\alpha Cr^2 - \frac{\alpha^2}{4}r^4 + S},$$
$$Z = \left(C - \frac{\alpha r^2}{2}\right)tg[\alpha(z+F)], \quad (29)$$

where $C$ and $S$ are constants of integration. Note, that equation $-\frac{\alpha^2}{4}r^4 + \alpha Cr^2 + S = 0$ has the roots $r_{1,2} = \pm\sqrt{2\frac{C}{\alpha} \pm \frac{2}{\alpha}\sqrt{C^2 + S}}$. Then, restrict our study by the case when $S = 0$, so

$$R = \pm\frac{1}{\cos[\alpha(z+F)]}\sqrt{\alpha\left(C - \frac{\alpha r^2}{4}\right)r^2},$$
$$Z = \left(C - \frac{\alpha r^2}{2}\right)tg[\alpha(z+F)]. \quad (30)$$

The mapping (30) transfers the region $0 \leq r < \sqrt{4C/\alpha}$, $(-\pi/2\alpha - F) < z < (\pi/2\alpha - F)$ into the whole space $(R, Z)$ with the exception of a disk in the plane $R = 0$ symmetrical about the axis $Z$. The disk radius is equal to $C$ (we can see this by fixing $z$ in (30) and considering $r$ as a parameter of the curves of equal potential).

Find the potential function $z = z(R, Z)$ which satisfies the Laplace equation. Let $\psi = z(R, Z)$ is the potential of a charge $Q$ of the equipotential conducting disk.

Expressing $r^2$ from the second equation (30) and substituting the obtained expression in the first equation (30), after simple transformations we obtain.

$$a^2 tg^4\left[\alpha(z+F)\right] + \left(a^2 - Z^2 - R^2\right) tg^2\left[\alpha(z+F)\right] - Z^2 = 0,$$

where $a^2 = C^2$. Then

$$\psi = z = -F + \frac{1}{\alpha} arctg \sqrt{\frac{Z^2 + R^2 - a^2 + \sqrt{\left(Z^2 + R^2 - a^2\right)^2 + 4a^2 Z^2}}{2a^2}} \qquad (31)$$

Far from the disk (when $Z \to \infty$) we put $\psi = 0$ and from (31) we obtain $F = \pi/2\alpha$. In addition, from the physical sense of the considered harmonic mapping (see above) it follows that the transformation conserves the charge located on the surface $S_Q = \pi(4C/\alpha) = 4\pi a/\alpha$ of the plane $z = const$. This charge is equal to $Q = \sigma S_Q = -\varepsilon_0 S_Q = -4\pi\varepsilon_0 a/\alpha$ and therefore $\alpha = -4\pi\varepsilon_0 a/Q$. Then

$$\psi = \frac{Q}{4\pi\varepsilon_0 a}\frac{\pi}{2} - \frac{Q}{4\pi\varepsilon_0 a} arctg \sqrt{\frac{Z^2 + R^2 - a^2 + \sqrt{\left(Z^2 + R^2 - a^2\right)^2 + 4a^2 Z^2}}{2a^2}} =$$

$$= \frac{Q}{4\pi\varepsilon_0 a} arctg \sqrt{\frac{2a^2}{Z^2 + R^2 - a^2 + \sqrt{\left(Z^2 + R^2 - a^2\right)^2 + 4a^2 Z^2}}}.$$

The obtained expression for the electrostatic potential of a thin conducting disc is identical to the expression obtained in [8] by another method.

b). Let us now return back to equations (28). Let us seek a solution in the form $Z = A(r) cth[\alpha(z+F)]$, $R = B(r)/sh[\alpha(z+F)]$. Then

$$R = -\frac{1}{sh[\alpha(z+F)]}\sqrt{-\frac{\alpha^2 r^4}{8} - \frac{\alpha r^2}{2}C + S},$$

$$Z = \left(\frac{\alpha r^2}{2} + C\right) cth[\alpha(z+F)],$$

(32)

where it is assumed that $\alpha < 0$, $C > 0$.

Consider only the case when $S = 0$, then

$$R = -\frac{1}{sh[\alpha(z+F)]}\sqrt{-\frac{\alpha r^2}{2}\left(\frac{\alpha r^2}{4} + C\right)},$$

$$Z = \left(\frac{\alpha r^2}{2} + C\right) cth[\alpha(z+F)].$$

(33)

The mapping (33) transfers the region $0 \leq r < \sqrt{-4C/\alpha}$, $0 < z < +\infty$ into the whole space $(R, Z)$ with the exception of an infinitely thin line segment (conductive segment) on the axis $Z$ with length $l = 2C$ (this can be seen graphically by fixing $z$ in (33) and considering $r$ as the parameter of the curves of equal potential).

Let us find the potential function $z = z(R, Z)$. Let $\psi = z(R, Z)$ is the potential of a charge $Q$ situated on an equipotential infinitely thin wire of finite length (conductive segment).

Expressing $r^2$ from the second equation (33) and substituting it in the first one, after simple transformations we find a solution

$$\psi = z = -F + \frac{1}{\alpha} arcth\sqrt{\frac{\left(2R^2 + Z^2 + \frac{l^2}{4}\right) - \sqrt{\left(2R^2 + Z^2 + \frac{l^2}{4}\right)^2 - Z^2 l^2}}{2Z^2}}.$$

Far from the conductive segment (when $Z \to \infty$) we have $\psi \to 0$ and therefore $F = 0$. In addition, from the physical sense of the considered harmonic mapping (see above) it follows that the transformation conserves the charge located on the surface $S_Q = \pi(-4C/\alpha) = -2\pi l/\alpha$ of the plane $z = const$. This charge is equal to $Q = \sigma S_Q = -\varepsilon_0 S_Q = 2\pi\varepsilon_0 l/\alpha$ and so, $\alpha = 2\pi\varepsilon_0 l/Q$. Then

$$\psi = \frac{Q}{2\pi\varepsilon_0 l} arcth \sqrt{\frac{\left(2R^2 + Z^2 + \frac{l^2}{4}\right) - \sqrt{\left(2R^2 + Z^2 + \frac{l^2}{4}\right)^2 - Z^2 l^2}}{2Z^2}} =$$

$$= \frac{Q}{2\pi\varepsilon_0 l} arcth \sqrt{\frac{1}{2} \frac{l^2}{\left(2R^2 + Z^2 + \frac{l^2}{4}\right) + \sqrt{\left(2R^2 + Z^2 + \frac{l^2}{4}\right)^2 - Z^2 l^2}}}.$$

Note that the surfaces of equal potential in this case are elongated ellipsoids of revolution, and for $\psi \to \infty$ these ellipsoids transform into an infinitely thin wire filament segment with length $l$. The fact that for a fixed charge $Q$ the potential $\psi \to \infty$ when $R \to 0$ and $Z \to 0$ physically means that an infinitely thin metal wire filament of finite length has infinitesimally small capacity.

**6. Harmonic mapping from the region with a orthogonal curvilinear coordinate system into the region with its orthogonal curvilinear coordinate system**

Consider in a space an electric field with a potential $\psi = f(x_3)$ depending only on the coordinate $x_3$ of the orthogonal curvilinear coordinate system $(x_1, x_2, x_3)$. Let the Lamé coefficients and orts in this coordinate system are $h_1$, $h_2$, $h_3$ and $\mathbf{i}_1$, $\mathbf{i}_2$, $\mathbf{i}_3$, respectively. Consider, by analogy with the previously discussed cases two nearby equipotential surfaces $a$ and $b$ defined by the values of potential $f(x_3)$ and $f(x_3 + dx_3)$, respectively. The mapping transfers these two surfaces into the equipotential surfaces $A$ and $B$. Consider the volume element $h_1 h_2 h_3 dx_1 dx_2 dx_3$ with the base area $ds = h_1 h_2 dx_1 dx_2$ on the equipotential plane $a$. The height of the element is equal to $h_3 dx_3$. We assume that $h_3 dx_3 = ds$. The image of this volume element in the space of images with orthogonal curvilinear coordinate system $(X_1, X_2, X_3)$ (with Lamé coefficients $H_1$, $H_2$, $H_3$ and orts $\mathbf{e}_1$, $\mathbf{e}_2$, $\mathbf{e}_3$) is the volume element with height

$$\left| \mathbf{e}_1 H_1 (X_1)'_{x_3} + \mathbf{e}_2 H_2 (X_2)'_{x_3} + \mathbf{e}_3 H_3 (X_3)'_{x_3} \right| dx_3$$

and base area

$$dS = \left|\left(\mathbf{e}_1 H_1(X_1)'_{x_1} + \mathbf{e}_2 H_2(X_2)'_{x_1} + \mathbf{e}_3 H_3(X_3)'_{x_1}\right) \times \left(\mathbf{e}_1 H_1(X_1)'_{x_2} + \mathbf{e}_2 H_2(X_2)'_{x_2} + \mathbf{e}_3 H_3(X_3)'_{x_2}\right)\right| dx_1 dx_2 =$$
$$= \left|\left(\mathbf{e}_1 H_1(X_1)'_{x_1} + \mathbf{e}_2 H_2(X_2)'_{x_1} + \mathbf{e}_3 H_3(X_3)'_{x_1}\right) \times \left(\mathbf{e}_1 H_1(X_1)'_{x_2} + \mathbf{e}_2 H_2(X_2)'_{x_2} + \mathbf{e}_3 H_3(X_3)'_{x_2}\right)\right| \frac{h_3 dx_3}{h_1 h_2}.$$

Then the harmonicity condition similar to (1) can be written as [8]:

$$\mathbf{e}_1 H_1(X_1)'_{x_3} + \mathbf{e}_2 H_2(X_2)'_{x_3} + \mathbf{e}_3 H_3(X_3)'_{x_3} = \frac{h_3}{h_1 h_2} \begin{vmatrix} \mathbf{e}_1 & \mathbf{e}_2 & \mathbf{e}_3 \\ H_1(X_1)'_{x_1} & H_2(X_2)'_{x_1} & H_3(X_3)'_{x_1} \\ H_1(X_1)'_{x_2} & H_2(X_2)'_{x_2} & H_3(X_3)'_{x_2} \end{vmatrix}. \quad (34)$$

In components, we obtain equations similar to the Cauchy-Riemann equations:

$$H_1(X_1)'_{x_3} = \frac{h_3}{h_1 h_2} \begin{vmatrix} H_2(X_2)'_{x_1} & H_3(X_3)'_{x_1} \\ H_2(X_2)'_{x_2} & H_3(X_3)'_{x_2} \end{vmatrix}, \quad (35)$$

$$H_2(X_2)'_{x_3} = -\frac{h_3}{h_1 h_2} \begin{vmatrix} H_1(X_1)'_{x_1} & H_3(X_3)'_{x_1} \\ H_1(X_1)'_{x_2} & H_3(X_3)'_{x_2} \end{vmatrix}, \quad (36)$$

$$H_3(X_3)'_{x_3} = \frac{h_3}{h_1 h_2} \begin{vmatrix} H_1(X_1)'_{x_1} & H_2(X_2)'_{x_1} \\ H_1(X_1)'_{x_2} & H_2(X_2)'_{x_2} \end{vmatrix}. \quad (37)$$

By differentiating the functions $x_1 = x_1(X_1, X_2, X_3)$, $x_2 = x_2(X_1, X_2, X_3)$ and $x_3 = x_3(X_1, X_2, X_3)$ we obtain

$$\begin{bmatrix} (X_1)'_{x_1} & (X_2)'_{x_1} & (X_3)'_{x_1} & 0 & 0 & 0 & 0 & 0 & 0 \\ (X_1)'_{x_2} & (X_2)'_{x_2} & (X_3)'_{x_2} & 0 & 0 & 0 & 0 & 0 & 0 \\ (X_1)'_{x_3} & (X_2)'_{x_3} & (X_3)'_{x_3} & 0 & 0 & 0 & 0 & 0 & 0 \\ 0 & 0 & 0 & (X_1)'_{x_1} & (X_2)'_{x_1} & (X_3)'_{x_1} & 0 & 0 & 0 \\ 0 & 0 & 0 & (X_1)'_{x_2} & (X_2)'_{x_2} & (X_3)'_{x_2} & 0 & 0 & 0 \\ 0 & 0 & 0 & (X_1)'_{x_3} & (X_2)'_{x_3} & (X_3)'_{x_3} & 0 & 0 & 0 \\ 0 & 0 & 0 & 0 & 0 & 0 & (X_1)'_{x_1} & (X_2)'_{x_1} & (X_3)'_{x_1} \\ 0 & 0 & 0 & 0 & 0 & 0 & (X_1)'_{x_2} & (X_2)'_{x_2} & (X_3)'_{x_2} \\ 0 & 0 & 0 & 0 & 0 & 0 & (X_1)'_{x_3} & (X_2)'_{x_3} & (X_3)'_{x_3} \end{bmatrix} \begin{bmatrix} (x_1)'_{X_1} \\ (x_1)'_{X_2} \\ (x_1)'_{X_3} \\ (x_2)'_{X_1} \\ (x_2)'_{X_2} \\ (x_2)'_{X_3} \\ (x_3)'_{X_1} \\ (x_3)'_{X_2} \\ (x_3)'_{X_3} \end{bmatrix} = \begin{bmatrix} 1 \\ 0 \\ 0 \\ 0 \\ 0 \\ 1 \\ 0 \\ 0 \\ 1 \end{bmatrix}$$

The determinant $J$ of this matrix is equal to

$$J = \begin{vmatrix} (X_1)'_{x_1} & (X_2)'_{x_1} & (X_3)'_{x_1} \\ (X_1)'_{x_2} & (X_2)'_{x_2} & (X_3)'_{x_2} \\ (X_1)'_{x_3} & (X_2)'_{x_3} & (X_3)'_{x_3} \end{vmatrix}^3 = D^3$$

where

$$D = \begin{vmatrix} (X_1)'_{x_1} & (X_2)'_{x_1} & (X_3)'_{x_1} \\ (X_1)'_{x_2} & (X_2)'_{x_2} & (X_3)'_{x_2} \\ (X_1)'_{x_3} & (X_2)'_{x_3} & (X_3)'_{x_3} \end{vmatrix} = (X_1)'_{x_3}\left((X_2)'_{x_1}(X_3)'_{x_2} - (X_3)'_{x_1}(X_2)'_{x_2}\right) +$$
$$+(X_2)'_{x_3}\left((X_1)'_{x_2}(X_3)'_{x_1} - (X_1)'_{x_1}(X_3)'_{x_2}\right) + (X_3)'_{x_3}\left((X_1)'_{x_1}(X_2)'_{x_2} - (X_1)'_{x_2}(X_2)'_{x_1}\right).$$

From (35)-(37) it follows

$$\frac{h_1 h_2}{h_3} \frac{H_1}{H_2 H_3}(X_1)'_{x_3} = (X_2)'_{x_1}(X_3)'_{x_2} - (X_2)'_{x_2}(X_3)'_{x_1}, \tag{38}$$

$$\frac{h_1 h_2}{h_3} \frac{H_2}{H_1 H_3}(X_2)'_{x_3} = (X_1)'_{x_2}(X_3)'_{x_1} - (X_1)'_{x_1}(X_3)'_{x_2}, \tag{39}$$

$$\frac{h_1 h_2}{h_3} \frac{H_3}{H_1 H_2}(X_3)'_{x_3} = (X_1)'_{x_1}(X_2)'_{x_2} - (X_1)'_{x_2}(X_2)'_{x_1}. \tag{40}$$

and therefore

$$D = \frac{h_1 h_2}{h_3} \left\{ \frac{H_1}{H_2 H_3} \left[ (X_1)'_{x_3} \right]^2 + \frac{H_2}{H_1 H_3} \left[ (X_2)'_{x_3} \right]^2 + \frac{H_3}{H_1 H_2} \left[ (X_3)'_{x_3} \right]^2 \right\}.$$

Consider that $D \neq 0$. From the equations (38)-(40) it follows

$$(x_1)'_{X_1} = D^{-1} \left( (X_2)'_{x_2} (X_3)'_{x_3} - (X_2)'_{x_3} (X_3)'_{x_2} \right), \tag{41}$$

$$(x_1)'_{X_2} = D^{-1} \left( (X_1)'_{x_3} (X_3)'_{x_2} - (X_1)'_{x_2} (X_3)'_{x_3} \right), \tag{42}$$

$$(x_1)'_{X_3} = D^{-1} \left( (X_1)'_{x_2} (X_2)'_{x_3} - (X_1)'_{x_3} (X_2)'_{x_2} \right); \tag{43}$$

$$(x_2)'_{X_1} = D^{-1} \left( (X_2)'_{x_3} (X_3)'_{x_1} - (X_2)'_{x_1} (X_3)'_{x_3} \right), \tag{44}$$

$$(x_2)'_{X_2} = D^{-1} \left( (X_1)'_{x_1} (X_3)'_{x_3} - (X_1)'_{x_3} (X_3)'_{x_1} \right), \tag{45}$$

$$(x_2)'_{X_3} = D^{-1} \left( (X_1)'_{x_3} (X_2)'_{x_1} - (X_1)'_{x_1} (X_2)'_{x_3} \right); \tag{46}$$

$$(x_3)'_{X_1} = D^{-1} \left( (X_2)'_{x_1} (X_3)'_{x_2} - (X_2)'_{x_2} (X_3)'_{x_1} \right), \tag{47}$$

$$(x_3)'_{X_2} = D^{-1} \left( (X_1)'_{x_2} (X_3)'_{x_1} - (X_1)'_{x_1} (X_3)'_{x_2} \right), \tag{48}$$

$$(x_3)'_{X_3} = D^{-1} \left( (X_1)'_{x_1} (X_2)'_{x_2} - (X_1)'_{x_2} (X_2)'_{x_1} \right); \tag{49}$$

By substituting the equations (47)-(49) into (38)-(40), we obtain

$$(x_3)'_{X_1} = \frac{h_1 h_2}{h_3 D} \frac{H_1}{H_2 H_3} (X_1)'_{x_3}, \quad (x_3)'_{X_2} = \frac{h_1 h_2}{h_3 D} \frac{H_2}{H_1 H_3} (X_2)'_{x_3}, \quad (x_3)'_{X_3} = \frac{h_1 h_2}{h_3 D} \frac{H_3}{H_1 H_2} (X_3)'_{x_3}, \tag{50}$$

or

$$(X_1)'_{x_3} = \frac{h_3 D}{h_1 h_2} \frac{H_2 H_3}{H_1} (x_3)'_{X_1}, \quad (X_2)'_{x_3} = \frac{h_3 D}{h_1 h_2} \frac{H_1 H_3}{H_2} (x_3)'_{X_2}, \quad (X_3)'_{x_3} = \frac{h_3 D}{h_1 h_2} \frac{H_1 H_2}{H_3} (x_3)'_{X_3}. \tag{51}$$

By substituting the expression (51) into (41)-(46), we obtain

$$(x_1)'_{X_1} = \frac{h_3}{h_1 h_2} \left( \frac{H_1 H_2}{H_3} (X_2)'_{x_2} (x_3)'_{X_3} - \frac{H_1 H_3}{H_2} (x_3)'_{X_2} (X_3)'_{x_2} \right),$$

$$(x_1)'_{X_2} = \frac{h_3}{h_1 h_2} \left( \frac{H_2 H_3}{H_1} (x_3)'_{X_1} (X_3)'_{x_2} - \frac{H_1 H_2}{H_3} (X_1)'_{x_2} (x_3)'_{X_3} \right),$$

$$(x_1)'_{X_3} = \frac{h_3}{h_1 h_2}\left(\frac{H_1 H_3}{H_2}(X_1)'_{x_2}(x_3)'_{X_2} - \frac{H_2 H_3}{H_1}(x_3)'_{X_1}(X_2)'_{x_2}\right);$$

$$(x_2)'_{X_1} = \frac{h_3}{h_1 h_2}\left(\frac{H_1 H_3}{H_2}(x_3)'_{X_2}(X_3)'_{x_1} - \frac{H_1 H_2}{H_3}(X_2)'_{x_1}(x_3)'_{X_3}\right),$$

$$(x_2)'_{X_2} = \frac{h_3}{h_1 h_2}\left(\frac{H_1 H_2}{H_3}(X_1)'_{x_1}(x_3)'_{X_3} - \frac{H_2 H_3}{H_1}(x_3)'_{X_1}(X_3)'_{x_1}\right),$$

$$(x_2)'_{X_3} = \frac{h_3}{h_1 h_2}\left(\frac{H_2 H_3}{H_1}(x_3)'_{X_1}(X_2)'_{x_1} - \frac{H_1 H_3}{H_2}(X_1)'_{x_1}(x_3)'_{X_2}\right).$$

Whence, after simple transformations, we obtain the expression

$$\left(\frac{H_1 H_3}{H_2}\right)(x_3)'_{X_2} = \left(\frac{h_1 h_2}{h_3}\right)\left((x_2)'_{X_1}(x_1)'_{X_3} - (x_1)'_{X_1}(x_2)'_{X_3}\right).$$

Similarly, we obtain expressions $\left(\frac{H_1 H_2}{H_3}\right)(x_3)'_{X_3} = \left(\frac{h_1 h_2}{h_3}\right)\left((x_1)'_{X_1}(x_2)'_{X_2} - (x_1)'_{X_2}(x_2)'_{X_1}\right)$ and

$$\left(\frac{H_2 H_3}{H_1}\right)(x_3)'_{X_1} = \left(\frac{h_1 h_2}{h_3}\right)\left((x_1)'_{X_2}(x_2)'_{X_3} - (x_1)'_{X_3}(x_2)'_{X_2}\right).$$

The result is the following condition equations for the inverse transformation, similar to the Cauchy-Riemann equations:

$$\begin{aligned}\left(\frac{H_2 H_3}{H_1}\right)(x_3)'_{X_1} &= \left(\frac{h_1 h_2}{h_3}\right)\left((x_1)'_{X_2}(x_2)'_{X_3} - (x_1)'_{X_3}(x_2)'_{X_2}\right), \\ \left(\frac{H_1 H_3}{H_2}\right)(x_3)'_{X_2} &= \left(\frac{h_1 h_2}{h_3}\right)\left((x_2)'_{X_1}(x_1)'_{X_3} - (x_1)'_{X_1}(x_2)'_{X_3}\right), \quad (52)\\ \left(\frac{H_1 H_2}{H_3}\right)(x_3)'_{X_3} &= \left(\frac{h_1 h_2}{h_3}\right)\left((x_1)'_{X_1}(x_2)'_{X_2} - (x_1)'_{X_2}(x_2)'_{X_1}\right).\end{aligned}$$

Consider a harmonic function $f(x_3)$ in the region $(x_1, x_2, x_3)$, that is let $\Delta_{(x_1,x_2,x_3)} f(x_3) = 0$.

This means that the function $f(x_3)$ satisfies the Laplace equation in this region:

$$\frac{1}{h_1 h_2 h_3}\left\{\frac{\partial}{\partial x_1}\left(\frac{h_2 h_3}{h_1}\frac{\partial}{\partial x_1}f(x_3)\right) + \frac{\partial}{\partial x_2}\left(\frac{h_3 h_1}{h_2}\frac{\partial}{\partial x_2}f(x_3)\right) + \frac{\partial}{\partial x_3}\left(\frac{h_1 h_2}{h_3}\frac{\partial}{\partial x_3}f(x_3)\right)\right\} = 0,$$

or

$$\frac{\partial}{\partial x_3}\left(\frac{h_1 h_2}{h_3}\frac{\partial}{\partial x_3}f(x_3)\right)=0.$$

Find the condition when the function $f(x_3) = f(x_3(X_1, X_2, X_3))$ satisfies the Laplace equation in a coordinate system $(X_1, X_2, X_3)$, i.e., that

$$\Delta_{(X_1,X_2,X_3)} f(x_3) =$$
$$= \frac{1}{H_1 H_2 H_3}\left\{\frac{\partial}{\partial X_1}\left(\frac{H_2 H_3}{H_1}\frac{\partial}{\partial X_1}f(x_3)\right)+\frac{\partial}{\partial X_2}\left(\frac{H_3 H_1}{H_2}\frac{\partial}{\partial X_2}f(x_3)\right)+\frac{\partial}{\partial X_3}\left(\frac{H_1 H_2}{H_3}\frac{\partial}{\partial X_3}f(x_3)\right)\right\} = 0$$

Indeed, using (52), we find

$$\Delta_{(X_1,X_2,X_3)} f(x_3) = \frac{1}{H_1 H_2 H_3}\left\{\frac{\partial}{\partial X_1}\left(\frac{H_2 H_3}{H_1} f'(x_3)\frac{\partial}{\partial X_1}x_3\right)+\frac{\partial}{\partial X_2}\left(\frac{H_3 H_1}{H_2} f'(x_3)\frac{\partial}{\partial X_2}x_3\right)\right.$$
$$\left.+\frac{\partial}{\partial X_3}\left(\frac{H_1 H_2}{H_3} f'(x_3)\frac{\partial}{\partial X_3}x_3\right)\right\} =$$
$$= \frac{1}{H_1 H_2 H_3}\left\{\frac{\partial}{\partial X_1}\left(\frac{h_1 h_2}{h_3} f'(x_3)\left((x_1)'_{X_2}(x_2)'_{X_3}-(x_1)'_{X_3}(x_2)'_{X_2}\right)\right)\right.$$
$$+\frac{\partial}{\partial X_2}\left(\frac{h_1 h_2}{h_3} f'(x_3)\left((x_2)'_{X_1}(x_1)'_{X_3}-(x_1)'_{X_1}(x_2)'_{X_3}\right)\right)$$
$$\left.+\frac{\partial}{\partial X_3}\left(\frac{h_1 h_2}{h_3} f'(x_3)\left((x_1)'_{X_1}(x_2)'_{X_2}-(x_1)'_{X_2}(x_2)'_{X_1}\right)\right)\right\}$$

Taking into account that

$$\frac{\partial}{\partial X_1}\left((x_1)'_{X_2}(x_2)'_{X_3}-(x_1)'_{X_3}(x_2)'_{X_2}\right)+\frac{\partial}{\partial X_2}\left((x_2)'_{X_1}(x_1)'_{X_3}-(x_1)'_{X_1}(x_2)'_{X_3}\right)+$$
$$+\frac{\partial}{\partial X_3}\left((x_1)'_{X_1}(x_2)'_{X_2}-(x_1)'_{X_2}(x_2)'_{X_1}\right)=0$$

we obtain

$$\Delta_{(X_1,X_2,X_3)} f(x_3) = \frac{1}{H_1 H_2 H_3}\left\{\left((x_1)'_{X_2}(x_2)'_{X_3}-(x_1)'_{X_3}(x_2)'_{X_2}\right)\frac{\partial}{\partial X_1}\left(\frac{h_1 h_2}{h_3} f'(x_3)\right)\right.$$
$$+\left((x_2)'_{X_1}(x_1)'_{X_3}-(x_1)'_{X_1}(x_2)'_{X_3}\right)\frac{\partial}{\partial X_2}\left(\frac{h_1 h_2}{h_3} f'(x_3)\right)$$
$$\left.+\left((x_1)'_{X_1}(x_2)'_{X_2}-(x_1)'_{X_2}(x_2)'_{X_1}\right)\frac{\partial}{\partial X_3}\left(\frac{h_1 h_2}{h_3} f'(x_3)\right)\right\}$$

We can see that $\Delta_{(X_1,X_2,X_3)} f(x_3) = 0$ when the following condition is satisfied

$$\left((x_1)'_{X_2}(x_2)'_{X_3} - (x_1)'_{X_3}(x_2)'_{X_2}\right)\frac{\partial}{\partial X_1}\left(\frac{h_1 h_2}{h_3}f'(x_3)\right) +$$

$$+ \left((x_2)'_{X_1}(x_1)'_{X_3} - (x_1)'_{X_1}(x_2)'_{X_3}\right)\frac{\partial}{\partial X_2}\left(\frac{h_1 h_2}{h_3}f'(x_3)\right) +$$

$$+ \left((x_1)'_{X_1}(x_2)'_{X_2} - (x_1)'_{X_2}(x_2)'_{X_1}\right)\frac{\partial}{\partial X_3}\left(\frac{h_1 h_2}{h_3}f'(x_3)\right) = 0.$$

Or at least the conditions

$$\frac{\partial}{\partial X_1}\left(\frac{h_1 h_2}{h_3}f'(x_3)\right) = 0, \quad \frac{\partial}{\partial X_2}\left(\frac{h_1 h_2}{h_3}f'(x_3)\right) = 0 \text{ и } \frac{\partial}{\partial X_3}\left(\frac{h_1 h_2}{h_3}f'(x_3)\right) = 0.$$

Based on the theorem just proved, consider two examples of finding the conditions of harmonic mappings for different systems of orthogonal curvilinear coordinates.

*Example № 1. The mapping of the region with a cylindrical coordinate system.*

Suppose we have a cylindrical coordinate system

$$(x_1, x_2, x_3) = (\varphi, z, r) \text{ and } h_1 = r, \ h_2 = h_3 = 1.$$

Let $f(x_3) = \ln r$, then

$$\frac{\partial}{\partial x_3}\left(\frac{h_1 h_2}{h_3}\frac{\partial}{\partial x_3}f(x_3)\right) = \frac{\partial}{\partial r}\left(r\frac{\partial}{\partial r}\ln r\right) = \frac{\partial}{\partial r}\left(r\frac{1}{r}\right) \equiv 0.$$

Consider the harmonic mapping from the space with the coordinate system $(x_1, x_2, x_3)$ into an image space with a coordinate system $(X_1, X_2, X_3)$. From the proved above it follows that the function $f(x_3) = \ln r = \ln x_3(X_1, X_2, X_3)$ satisfies the Laplace equation in some coordinate system $(X_1, X_2, X_3)$. Now suppose that in the image space, there is a cylindrical coordinate system

$$(X_1, X_2, X_3) = (R, \Phi, Z) \text{ and } H_1 = 1, \ H_2 = R \ H_3 = 1.$$

Then the equations similar to Cauchy-Riemann (38) - (40) take the form

$$\frac{r}{R}R'_r = \Phi'_\varphi Z'_z - \Phi'_z Z'_\varphi,$$

$$rR\Phi'_r = R'_z Z'_\varphi - R'_\varphi Z'_z,$$

$$\frac{r}{R}Z'_r = R'_\varphi \Phi'_z - R'_z \Phi'_\varphi.$$

*Example № 2. The mapping of the region with a spherical coordinate system.*

Suppose we have a spherical coordinate system

$$(x_1, x_2, x_3) = (\theta, \varphi, r) \text{ and } h_1 = r, \ h_2 = r\sin\theta, \ h_3 = 1.$$

Let $f(x_3) = \dfrac{1}{r}$. This function satisfies the Laplace equation in the space of preimages $(x_1, x_2, x_3)$, as

$$\frac{\partial}{\partial x_3}\left(\frac{h_1 h_2}{h_3}\frac{\partial}{\partial x_3}f(x_3)\right) = \frac{\partial}{\partial r}\left(\frac{r \cdot r\sin\theta}{1}\frac{\partial}{\partial r}\left(\frac{1}{r}\right)\right) = -\frac{\partial}{\partial r}(\sin\theta) \equiv 0$$

Consider the harmonic mapping from the space with the coordinate system $(x_1, x_2, x_3)$ into an image space with a coordinate system $(X_1, X_2, X_3)$. From the proved above it follows that the function $f(x_3) = 1/r = 1/x_3$ satisfies the Laplace equation in the space of images $(X_1, X_2, X_3)$.

Suppose that in the image space there is a cylindrical coordinate system such that $(X_1, X_2, X_3) = (R, \Phi, Z)$ and $H_1 = 1$, $H_2 = R$, $H_3 = 1$. Then the equations similar to Cauchy-Riemann (38) - (40) take the form

$$r^2 \sin\theta \frac{1}{R} R'_r = \Phi'_\theta Z'_\varphi - \Phi'_\varphi Z'_\theta,$$

$$r^2 \sin\theta R\Phi'_r = R'_\varphi Z'_\theta - R'_\theta Z'_\varphi,$$

$$r^2 \sin\theta \frac{1}{R} Z'_r = R'_\theta \Phi'_\varphi - R'_\varphi \Phi'_\theta.$$

## 7. Conclusion

A new method for obtaining harmonic mappings has been found. Its effectiveness for the calculation of electrostatic fields has been demonstrated. The method is based on a physically clear concept of local charge conservation under the harmonic mappings. The obtained transformation formulas can be regarded as a system of equations, which produces three-dimensional harmonic fields in domains with complex geometry.

# References

[1] Lavrentiev M A 1946 *Conformal mappings* (Moscow-Leningrad: Gostehizdat) (in Russian).

[2] Lavrentiev M A 1951 Dirichlet problem for a narrow layer Trudi matem. in-ta AS USSR **38**, 146 (in Russian).

[3] Sterensky L N 1977 *Theory of wave motions of fluid* (Moscow: Nauka) (in Russian).

[4] Maxwell J C 1989 *A treatise on electricity and magnetism* V. I, II (Moscow: Nauka).

[5] Gergen J J 1930 Mapping of a general type of three – dimensional region on a sphere *Amer. J. of Mathem.* **52** 197.

[6] Petrin A B, VINITI, Moscow, Dep. 03.02.84, №640-84, p.24-27 (1984) (in Russian). (The copies of the paper may be ordered by POD mail at the following address: Department of Scientific Papers Deposition, VINITI, Usievicha Street 20, Moscow 125219, Russia).

[7] Yanushkauskas A I 1982 *Three-dimensional analogues of conformal mappings* (Novosibirsk: Nauka) (in Russian).

[8] Landau L D and E. M. Lifshitz E M 1984 *Course of Theoretical Physics, Vol. 8: Electrodynamics of Continuous Media*, 2nd ed. (Oxford: Pergamon).